 \documentclass [12pt,a4paper      ]{article}
\usepackage{times}

\DeclareFontFamily{OT1}{times}{}
\DeclareFontShape {OT1}{times}{m }{n }{ <-> ptmr }{}
\DeclareFontShape {OT1}{times}{bx}{n }{ <-> ptmb }{}
\DeclareFontShape {OT1}{times}{m }{it}{ <-> ptmri}{}
\DeclareFontShape {OT1}{times}{bx}{it}{ <-> ptmbi}{}
\usepackage{amsmath}
\usepackage{amsfonts}
\usepackage{amssymb}
\usepackage{latexsym}
\newcommand{\cl}{C \kern -0.1em \ell} 

\newcommand{\CON}{\overline}          

\newcommand{\VEC}{\vec{\kern +.1em[}} 
\newcommand{\TOR}{\vec{\kern +.2em]}} 
\newcommand{\BRA}{\langle\kern -.2em\langle} 
\newcommand{\KET}{\rangle\kern -.2em\rangle} 

\newcommand{\DUA}{\widetilde}         

\numberwithin{equation}{section}
\begin{document}

\title{\bf\vspace{-2.5cm} The Conservation Laws in the Field Theoretical
                          Representation of Dirac's 
                          Theory\footnote{\emph{Editorial note}: Published
                          in Zeits.\ f.\ Phys.\ {\bf 57} (1929) 484--493,
                          reprinted and translated in~\cite{LAN1929D}.
                          This is Nb.~3 in a series of four papers
                          on relativistic quantum mechanics
                          \cite{LAN1929B,LAN1929C,LAN1929D,LAN1930A}
                          which are extensively discussed
                          in a commentary by Andre Gsponer
                          and Jean-Pierre Hurni \cite{GSPON1998B}.
                          Initial translation by J\'osef Illy and 
                          Judith Konst\'ag Mask\'o. Final translation
                          and editorial notes by Andre Gsponer. }}

\author{By Cornel Lanczos in Berlin\\ (Received on August 13, 1929)}

\date{Version ISRI-04-12.3 ~~ \today}

\maketitle

\begin{abstract}

We show that in the new description, Dirac's ``current vector'' is not related to a vector but to a tensor: the ``stress-energy tensor.'' Corresponding to Dirac's conservation law, we have the conservation laws of momentum and energy. The stress-energy tensor consists of two parts: an ``electromagnetic'' part, which has the same structure as the stress-energy tensor of the Maxwell theory, and a ``mechanical'' part, as suggested by hydrodynamics. The connection between these two tensors, which appears organically here, eliminates the well-known contradictions inherent in the dynamics of electron theory. (\emph{Editorial note:} In this paper Lanczos continues to discuss his ``fundamental equation,'' from which he consistently derives Proca's equation and its stress-energy tensor.)

\end{abstract}

In two previous papers,\footnote{Zeits.\ f.\ Phys.\ {\bf 57} (1929) 447 and 474, 1929. (\emph{Editorial note:} Refs.~\cite{LAN1929B} and ~\cite{LAN1929C}.)} the author proposed a new way of describing Dirac's theory; namely, exclusively on the basis of the normal relativistic space-time structure and operating with customary field theoretical concepts only. In one respect, the new description displayed a peculiar deficiency: no vector could be found that would correspond to the fundamental zero-divergence ``current vector'' of Dirac's theory. Namely, the vector which could be considered as a form analogous to Dirac's current vector [cf., expression (90) in the first paper], is \emph{not} divergence free, whereas the formation which is really divergence free [cf., expression (13) in the second paper] does not represent a vector. This difficulty is solved --- as the author realized in the meantime --- by a fact which allows a much larger perspective for the field theoretical description and seems to confirm the inherent validity of the whole development to a great extent.

   As we mentioned before, the two Dirac equations for $H$ and $H'$ [see equation (8) in the second paper] are completely equivalent to our whole system of field equations. Hence, we can form the ``current vector'' for these two Dirac equations (which is not complex and therefore actually represents only one vector) and in this way we obtain two zero-divergence expressions. We should not call these ``vectors'' because the vector character disappeared with our transformation properties of $F$ and $G$. However, the zero divergence follows simply from the field equations and is independent of the transformation properties.

   Hence, we have two zero divergence quaternions:
\begin{gather*}\label{1}
H \CON{H}^* , \quad H' \CON{H}'{}^*
\tag{1}
\end{gather*}
and it is obvious that any arbitrary linear combination of them will be divergence-free as well.

If we write for a moment:
\begin{gather*}\label{2}
A = \frac{1}{2} (F + G) , \quad B = \frac{1}{2} (F - G) ,
\tag{2}
\end{gather*}
then:
\begin{equation*}
\left.
\begin{aligned}\label{3}
H & = A + i B j_z, \\
H' & = A - i B j_z. 
\end{aligned} 
\right\} 
\tag{3}
\end{equation*}
The zero divergence property holds for the following two quaternions as well:
\begin{equation*}\label{4}
\left.
\begin{aligned}
A \CON{A}^* + B \CON{B}^* & = \frac{1}{2} (F\CON{F}^* + G \CON{G}^* )  ,\\
B j_z \CON{A}^* + A j_z \CON{B}^* & = \frac{1}{2} (F j_z \CON{F}^* - G j_z \CON{G}^* ) .
\end{aligned} 
\right\}
\tag{4}
\end{equation*}

   It is obvious, however, that the quaternion unit $j_z$ cannot be distinguished from the remaining spatial units. The choice of $j_z$ was only due to a special way of writing down the Dirac equation, which thereby requires a special ordering of the $\psi$-quantities. Accordingly, we may use either $j_x$ or $j_y$ instead of $j_z$.

   In this way, we obtain four divergence-free quaternions, which we can write down in the following compact form:
\begin{equation*}\label{5}
F j_{\alpha} \CON{F}^* + G \CON{j}_{\alpha} \CON{G}^* ,
\tag{5}
\end{equation*}
where $ j_{ \alpha} $ stands for one of the four quaternion units.

   The fact that the Dirac divergence equation is quadrupled here suggests that we have a vectorial divergence instead of a scalar one. If this is true, then the set of the four quaternions \eqref{5} should be equivalent to a tensor. In actual fact, this is the case.

   By means of a vector $V$, one can namely form a vector again from an antisymmetric tensor F by means of the following quaternion product:
\begin{equation*}\label{6}
F V \CON{F}^* .
\tag{6}
\end{equation*}
Indeed,
\begin{equation*}\label{7}
F' V' \CON{F}'{}^* = p F \CON{p} p V \CON{p}^* p^* \CON{F}^* \CON{p}^* = p (F V \CON{F}^* ) \CON{p}^* .
\tag{7}
\end{equation*}
However, for \eqref{6} we can write:
\begin{equation*}\label{8}
F V \CON{F}^* = (F j_{\nu} \CON{F}^*) V_{\nu} ,
\tag{8}
\end{equation*}
and the vector character of a quaternion $Q$ can also be expressed by saying:
\begin{equation*}\label{9}
Q_{\mu} U_{\mu} = \text{invariant} ,
\tag{9}
\end{equation*}
where the components of quaternions $Q$ are denoted by $Q_{\mu}$ and where $U$ is a vector. Hence we have:
\begin{equation*}\label{10}
(F j_{\nu} \CON{F}^*)_{\mu} U_{\mu} V_{\nu} = \text{invariant} ,
\tag{10}
\end{equation*}
and this means, according to the definition of a tensor, that the quantities:
\begin{equation*}\label{11}
(F j_i \CON{F}^*)_k ,
\tag{11}
\end{equation*}
are tensor components. In other words: if we write the components of each quaternion $F j_{ \alpha} \CON{F}^*$ one after the other each in a line, then the four lines taken together yield an array of 16 quantities which are tensor components.

   Something analogous can be done with the vector $G$. There we can form a vector by means of the product $G \CON{V} G$ or even by means of:
\begin{equation*}\label{12}
G \CON{V} \, \CON{G}^* = (G \CON{j}_{\nu} \CON{G}^*) V_{\nu} .
\tag{12}
\end{equation*}
Then it holds that:
\begin{equation*}\label{13}
G' \CON{V}' \CON{G}'{}^* = p G \CON{p}^* p^* \CON{V} \CON{p} p \CON{G}^* \CON{p}^* = p (G \CON{V} \, \CON{G}^* ) \CON{p}^* ,
\tag{13}
\end{equation*}
is valid. That is, the quantities:
\begin{equation*}\label{14}
(G \CON{j}_i \CON{G}^*)_k ,
\tag{14}
\end{equation*}
also form the components of a tensor. It is expedient to apply a factor of $-\frac{1}{2}$ and therefore we put:
\begin{equation*}\label{15}
T_{ik}  = -\frac{1}{2} (F j_i \CON{F}^*)_k , 
\tag{15}
\end{equation*}
and
\begin{equation*} \label{16}
U_{ik}  = -\frac{1}{2} (G \CON{j}_i \CON{G}^*)_k . 
\tag{16}
\end{equation*}
The zero divergence tensor, which we shall denote by $W_{ik}$, is composed of these two tensors:\footnote{The letter $W$ should not remind one of probability (``Wahrscheinlichkeit''). If the Dirac vector could be interpreted as a ``probability flux'' (``Wahrscheinlichkeitsfluss''), then an analog interpretation for a tensor of second-order, here replacing the Dirac vector, would hardly have any meaning. Therefore, I think that at this stage no compromise is any longer possible between the ``reactionary'' viewpoint represented here (which aims at a complete field theoretical description based on the normal space-time structure) and the probability theoretical (statistical) approach.}
\begin{equation*}\label{17}
W_{ik} = T_{ik} + U_{ik} .
\tag{17}
\end{equation*}

   Thus, Dirac's conservation law for the four quaternions \eqref{5} appears in the form of a divergence equation for this tensor:
\begin{equation*}\label{18}
\text{div} (W_{ik}) = \frac{\partial W_{i \nu}}{\partial x_{\nu}} = 0 ,
\tag{18}
\end{equation*}
which describes the conservation laws of momentum and energy.

   In actual fact, the tensor $W_{ik}$ occurring here, whose divergence vanishes, can really with good reason be called a ``stress-energy tensor'' and thus we arrive at the following remarkable result:

   In place of the Dirac current vector the stress-energy tensor occurs, and in place of the Dirac conservation law the momentum-energy law occurs.

   The Dirac current vector was an extension for those scalars $\psi\psi^*$ interpreted by Schr\"odinger as ``the density of electricity.'' Here the same vector will be extended by one more rank: to a tensor of second order.\footnote{This procedure resembles the development of gravitation theory where Newton's scalar potential was extended to a tensor of second-order by Einstein.} However, the larger manifold of quantities may well be taken into account if we think of the fundamental significance the stress-energy tensor has for dynamics and of the fundamental significance of the Riemannian curvature tensor.  Thereby, one can presume a metrical background for the whole theory proposed here, as well as a hidden connection with the most important and far-reaching branch of physics: with the general theory of relativity.

   Strangely, the stress-energy tensor given by \eqref{17} is not symmetric.

   Let us first consider the tensor \eqref{15}. It can be written in vector analytical terms as follows:
\begin{equation*}\label{15'}
T_{ik} = S F_{ik} - M \DUA{F}_{ik} - \frac{1}{2} (S^2 + M^2) g_{ik} + (F_i^{\nu} F_{k \nu} - \frac{1}{4} F_{\mu \nu} F^{\mu \nu} g_{ik}) .
\tag{15'}
\end{equation*}
The first two terms are antisymmetric, the others are symmetric. 

The other tensor \eqref{16} appears in the form:
\begin{equation*}\label{16'}
U_{ik} = \DUA{(M_i S_k - M_k S_i)} + M_i M_k + S_i S_k - \frac{1}{2} (M_{\nu} M^{\nu} + S_{\nu} S^{\nu}) g_{ik} .
\tag{16'}
\end{equation*}
Here, too, an antisymmetric term is produced by the interaction between the two vectors $S_i$ and $M_i$.

   It is quite remarkable that the stress-energy tensor becomes symmetric when all those quantities which are extraneous to the Maxwell theory drop out, that is, if we set equal to zero the scalars $S$ and $M$, as well as the magnetic vector $M_i$. In the first paper we indicated --- without an external field --- that this constraint is really possible, whereas in the second paper we saw that the same was not feasible after introducing the vector potential. Here it is indicated again that the introduction of the vector potential in the equations was not performed in the right way. In fact, the fundamental meaning of the stress-energy tensor would be lost if we sacrificed its symmetry --- there is no doubt about that.

   If we retain only the electromagnetic field strength $F_{ik}$ and the electric current vector $S_i$ as fundamental quantities, then the now symmetric stress-energy tensor appears to be composed of two parts.

   The first ``electromagnetic'' part $T_{ik}$ is fully identical with the Maxwell stress-energy tensor of the electromagnetic field:
\begin{equation*}\label{19}
T_{ik} = F_i^{\nu} F_{k \nu} - \frac{1}{4} F_{\mu \nu} F^{\mu \nu} g_{ik} .
\tag{19}
\end{equation*}

   The second part $U_{ik}$ can also be given a certain meaning, in view of a similar formulation in mechanics. It is:
\begin{equation*}\label{20}
U_{ik} = S_i S_k - \frac{1}{2} S_{\nu} S^{\nu} g_{ik} .
\tag{20}
\end{equation*}
This tensor can be regarded as a ``mechanical'' stress-energy tensor. In fact, the symmetric tensor:
\begin{equation*}\label{21}
\mu_0 u_i u_k ,
\tag{21}
\end{equation*}
is introduced in the field theoretical description of the mechanical momentum-energy current as a ``kinetic momentum-energy tensor'' (Minkowski).\footnote{Cf., e.g., W. Pauli, Theory of Relativity (Teubner, Leipzig and Berlin, 1921) p.~675; M. v. Laue, The theory of Relativity, Part 1, 4th Edition (Friedr. Vieweg \& Sohn, Braunsweig, 1921) p.~207.} Here $\mu_0 $ is the scalar mass density and $u_i$ is the velocity vector. The term $S_i S_k$ is obviously quite analogous, except that the current vector is replaced by the velocity. And this analogy goes further if we take into consideration that, in a static spherically symmetric solution, the average value of the spatial components of $S_i$ will necessarily vanish and only a time part can remain. This means that the average value of the vector $S_i$ points in the direction of the velocity indeed. (In a system at rest, the latter has only one time component.)

   The second term of \eqref{20} is also well-known from hydrodynamics.  There, an additional term $p g_{ik}$ appears for the matter tensor if $p$ means the hydrostatic pressure (which is a scalar). The relation \eqref{20} indicates that the mechanical mass density $\mu_0 $ is accompanied by a hydrostatic pressure of the value $\mu_0 /2 $. This pressure is extremely high if we consider that in the CGS system we have to multiply by $c^2$. For water we would obtain the enormous amount of $4.5 \times 10^{14}$ atm!\footnote{This remarkable result resembles the well-known conclusion of the theory of relativity that each mass $m$ is connected with the enormous amount of energy $mc^2$. Similarly to the kinetic energy in mechanics representing a small difference contribution compared to the rest energy, the common hydrostatic pressure of gravitational origin appears here as a second-order quantity compared to the enormous ``eigenpressure'' of matter. This pressure proves to be positive indeed --- i.e., it is directed inwards, in spite of the seemingly opposite sign of the last tern in equation \eqref{20}. For the square of the length of the velocity vector $u^k = i dx^k/ds$  is not $+1$ but $-1$.} However, this pressure is not meant macroscopically for neutral materials. We should consider it rather as the ``cohesion pressure'' required for the construction of an electron, i.e., to compensate for the strong electric repulsive force.\footnote{Although the electron is, of course, ``smeared,'' an estimate of the dimensions may be of interest for a comparison with electron theory. Let us consider a spherical shell of radius $a$ with evenly distributed charge and mass. Then the hydrostatic pressure connected with the mass density ${M}/{4 \pi \alpha^2}$ implied an inwards directed force of ${M c^2}/{4 \pi \alpha^3}$ per unit surface. The outwards directed electric pulling force amounts to $\frac{1}{2} \left({e}/{4 \pi \alpha^2} \right)^2$. The balance between the two forces requires that:
\begin{equation*}
\frac{M c^2}{4 \pi a^3} = \frac{1}{2} \left( \frac{e}{4 \pi a^2} \right)^2 ,
\end{equation*}
From this result, we obtain for the mechanical mass:
\begin{equation*}
M = \frac{e^2}{8 \pi c^2 a} ,
\end{equation*}
Just as large is the field's electrostatic energy divided by $c^2$, i.e., the ``electromagnetic mass'' calculated from electron theory.}

   If we consider that the divergence of the Maxwell tensor yields the Lorentz force and the divergence of the mechanical tensor the inertia force, then we can see on these grounds how the dynamics of the electron follows as a harmonious closed whole, which has never been possible on the basis of classical field theory. Indeed, though one had probably guessed that electromagnetic quantities needed to be completed by mechanical ones so that they can supply the ``cohesion forces,'' i.e., prevent the electron from exploding into pieces and permit a differential formulation of dynamics. However, there had been no basis for expecting an organic merging of mechanics and electrodynamics.

   The field equations obtained here provide such an inherent connection on account of the double coupling between field strength and current vector, The current vector ceases to be a ``material'' quantity forced from outside which does not really belong to the field and is only meant to avoid a singularity. Rather, here it represents an actual field quantity which is determined by the field equations. Similarly, the zero divergence of the matter tensor is no longer a heuristic principle for obtaining the dynamics in addition to the field equations, but these basic dynamic equations appear as a necessary consequence of the field equations. Thus, the inner closure is of the same structure as in the theory of general relativity where the divergence equation describes a mathematical identity of the curvature tensor and the principle of geodesics is affirmed already by the field equations.\footnote{Though it seems plausible to place electron dynamics on this basis, our approach is not yet sufficient for this. Namely, the divergence equation as a mathematical consequence of the field equations does not contain anything which would go beyond this. However, the field equations are linear and permit, therefore, the superposition principle which a priori excludes a dynamic influence. This discrepancy is most probably connected with the already often-mentioned difficulty: with the incorporation of the vector potential into the equations. This was first done on the basis of the quantum mechanical rule but it led to obviously unsatisfactory results. Such an incorporation would not at all be necessary since the field quantities are obviously available already in a sufficient choice and especially the current vector already plays the role of the vector potential in the ``feedback,'' so this should not be introduced separately as an extraneous element. The extension of the equations by the vector potential appears in this approach only as a makeshift for a not yet known nonlinearity of the system. Then the divergence equation could really contain the motion principle without becoming incompatible with the superposition principle (which is then not valid any more).\label{page490}}

   Thus, the connection \eqref{17} of the two essentially different tensors \eqref{15} and \eqref{16} is not an extraneous one but is unequivocally determined by the structure of the theory. For neither the one (``electromagnetic'') nor the other (``mechanical'') part has a vanishing divergence but only the given sum whereby no factor and no sign remain free.  Essential difficulties and inherent contradictions of electron theory are thus eliminated and the relationship revealed here, unexpectedly, is so impressive that it can hardly be doubted that this way leads us to deeper knowledge. Of course, we are not yet able to solve the electron problem by this alone, for obviously there are still essential features missing. On the one hand, this is not at all conceivable on the basis of a linear system of equations and, on the other hand, these equations (just like the classical field equations) do not have regular ``eigensolutions'' of the kind which would give stationary energy nodes --- as would be expected for a really satisfactory ``theory of matter.''

   However, the major objection which can be made to the conjecture, that quantum theory in the end would lead to a correction of classical field theory through the here revealed connection between Dirac's theory and the Maxwell equations, is that it does not yield the classical theory of the electron even as a ``first approximation.'' To perform a comparison with electron theory, we should once more write down the reduced equation system (98) of the first paper which we obtained as a final result for the free electron if we omit all quantities which are extraneous to the theory of the electron. As the only difference, we shall introduce another field quantity:
\begin{equation*}\label{22}
\varphi_i = \frac{S_i}{\alpha} ,
\tag{22}
\end{equation*}
instead of the current vector $S_i$. We shall call this quantity the ``vector potential'' in accordance with the feedback equation.

   Then we have the equations:
\begin{equation*}\label{23}
\left.
\begin{aligned}
\frac{\partial F_{i \nu}}{\partial x_{\nu}} & = \alpha^2 \varphi_i , \\
\frac{\partial \varphi_i}{\partial x_k} - \frac{\partial \varphi_k}{\partial x_i} & = F_{ik} .
\end{aligned}
\right\}
\tag{23}
\end{equation*}
In the vacuum equations of the electron theory, the right-hand side term of the first equation is missing.\footnote{\emph{Editorial note:} Eq.~\eqref{23} is the correct wave-equation for a massive spin~1 particle, and  Eq.~\eqref{17} the corresponding stress-energy tensor, both to be rediscovered by Proca in 1936.  For more details, see section 11 in \cite{GSPON1998B}.}  Thus, we obtain the classical field equations if we let the constant $\alpha$ converge to~0. However, since the constant $\alpha = {2 \pi m c}/{h}$ contains the Planck constant $h$ in the numerator, this limiting process does not means $h \rightarrow 0$ but $h \rightarrow \infty$. The macroscopic behaviors of the electron will thus be characterized by the unnatural transition $h \rightarrow \infty$ instead of the expected transition $h \rightarrow 0$. In fact, one could consider the electron theory as a first approximation only if the constant of the theory was very small. In actual fact, this constant is very large: $\alpha = 2.59 \times 10^{10} \, \text{cm}^{-1}$. That is, even if the equation were already completed by (still unknown) quadratic terms --- which is required anyway (cf., footnote \ref{page490})
 --- this would not yet solve the problem that the macrocospic behavior of the electron is certainly incorrectly described.  Namely, at larger distances, where the quadratic terms have already been reduced to zero and the linear approximation is justified, one would not obtain potentials decreasing with $1/r$, the potential behavior would rather be characterized by ${ e^{-\alpha r} }/{r} $.  Here $\alpha$ is the strong attenuation constant which definitely contradicts experience since it completely excludes the action of an electron over a large distance.

   If we consider it plausible that the quantum mechanical reaction of the single electron (spin action, etc.) acts over very short distances, then we can also say the following: At short distances the free electron behaves as if the constant $\alpha$ were very large and at long distances as if the same constant were very small. Unless we want to accept the highly unlikely dualism that there are also special ``quantum mechanical'' processes in addition to the customary field theoretical ones, we necessarily arrive at the requirement that the constant $\alpha^2$ of our theory should not be considered as an actual constant.  It should be considered as a field function which depends on the fundamental field quantities themselves in some still unknown way.\footnote{The supposition of a correlation with the scalar Riemannian curvature (which also has the dimension $\text{cm}^{-2}$) is hardly rejectable in view of Einstein's theory of gravitation. \emph{Editorial note:}  A nonlinear generalization of the present theory in which the constant $\alpha^2$ is replaced by $\alpha^2\sigma$ where $\sigma$ is a field function is considered in the last paper in this series, i.e., Ref.~\cite{LAN1930A}.}

   Then everything would fall into place. Then the term with $\alpha^2$ is no longer a linear term but one of higher order. For the linear approximation, we would obtain the vacuum equations of the classical electron theory. The $\alpha^2$ function would practically decrease to zero in the peripheral range, whereas it would be expected to have a practically smooth functional form of the given order of magnitude in the central domain, i.e., in the immediate vicinity of the electron's center. Then one could understand why the de Broglie-Schr\"odinger wave equation with constant $\alpha$ cannot characterize a single electron but only a large ``swarm'' of electrons. Then statistical averages over multiple spatial neighborhoods of different electrons could result in a functional form of $\alpha$ which is sufficiently constant for a larger domain, whereas $\alpha$ decreases to zero very rapidly for a single electron.

   In a comprehensive field theory, it would hardly be conceivable to introduce a quantity as a ``universal constant'' which contains the mass of the electron. It would then be hopeless to understand the mass difference between electron and proton.

   Of course, this new hypothesis would also influence the matter tensor. In fact, the constant $\alpha$ appears in the mechanical part of the matter tensor, and the vector potential according to \eqref{22} is introduced instead of the current vector. Then we have:
\begin{equation*}\label{24}
U_{ik} = \alpha^2 (\varphi_i \varphi_k - \frac{1}{2} \varphi_{\nu} \varphi^{\nu} g_{ik} ) .
\tag{24}
\end{equation*}
In the peripheral domain, where $\alpha$ has practically become zero, the mechanical component drops out and only the customary electromagnetic stress-energy tensor remains. However, in the central domain (just where it is demanded for the construction of the electron!) the strong mechanical component and, especially, the high cohesion pressure becomes active. Of course, the expression \eqref{24} and the sum to be formed from it on the basis of \eqref{17} can be considered for the matter tensor only in first approximation, with slowly varying $\alpha$, because the zero divergence of this tensor was proven by assuming a constant $\alpha$.

   If the above anticipated possibilities should prove really viable, quantum mechanics would cease to be an independent theory.  It would merge with a deeper ``theory of matter,'' which relies on regular solutions of nonlinear differential equations --- in the final connection, it would be absorbed into the ``world equations'' of the Universe. Then the ``matter-field'' dualism would become just as obsolete as the ``particle-wave'' dualism.

Berlin-Nikolassee, August 1929.

\end{document}